\documentclass[useAMS,usenatbib]{mn2e}
\usepackage{epsfig}
\usepackage{rotate}
\usepackage{amssymb}
\usepackage{amsmath}
\usepackage{amstext}
\newcommand{\kms}{km\,s$^{-1}$}
\newcommand{\chandra}{{\it Chandra}}

\newcommand{\xper}{$\xi$~Per}

\newcommand{\ha}{${\rm H}{\alpha}$}

\title[CIR Modulation of X-rays]{CIR Modulation of the X-ray Flux from 
the O7.5 III(n)((f)) Star $\xi$~Persei\thanks{Based on data obtained with 
the Chandra X-Ray Observatory.}?}
\author[D. Massa et al.]
{\parbox{\textwidth}{D. Massa$^{1}$\thanks{E-mail: dmassa@spacescience.org}, 
L. Oskinova$^{2}$, A. W. Fullerton$^{3}$, R. K. Prinja$^{4}$, D. A. 
Bohlender$^{5}$, N. D. Morrison$^{6}$, M. Blake$^{7}$ and W. Pych$^{8}$}
\vspace{0.4cm}\\
$^{1}$Space Science Institute, 4750 Walnut St, Suite 205, Boulder, CO 
      80301, USA\\
$^{2}$Institute for Physics and Astronomy, University Potsdam, 14476 
      Potsdam, Germany\\
$^{3}$Space Telescope Science Institute, 3700 San Martin Dr., Baltimore, MD 
      21218, USA\\
$^{4}$Dept. of Physics \& Astronomy, University College London, Gower 
      St., London, England WC1E 6BT\\
$^{5}$National Research Council of Canada, 5071 W. Saanich Rd., Victoria, 
      BC V9E 2E7, Canada\\
$^{6}$Ritter Observatory, Dept. of Physics \& Astronomy, The University of 
      Toledo, Toledo, OH 43606-3390, USA\\
$^{7}$UNA Box 5263 University of North Alabama Florence, Alabama 35632\\
$^{8}$Nicolaus Copernicus Astronomical Centre Polish Academy of Sciences
  ul. Bartycka 18 00-716 Warszawa Poland
      }
\begin{document}

\date{Accepted 1988 December 15. Received 1988 December 14; in original 
      form 1988 October 11}

\pagerange{\pageref{firstpage}--\pageref{lastpage}} \pubyear{2002}

\maketitle

\label{firstpage}

\begin{abstract}
We analyze a 162 ks HETG \chandra\ observation of the O7.5 III(n)((f)) 
star $\xi$~Per, together with contemporaneous H$\alpha$ observations.
The X-ray spectrum of this star is similar to other single O stars, and not 
pathological in any way.  Its UV wind lines are known to display cyclical 
time variability, with a period of 2.086~days, which is thought to be 
associated with co-rotating interaction regions (CIRs).  We examine the 
\chandra\/ and \ha\ data for variability on this time scale.  We find that 
the X-rays vary by $\sim 15$\% over the course of the observations and that 
this variability is out of phase with variable absorption on the blue wing 
of the \ha\ profiles (assumed to be a surrogate for the UV absorption 
associated with CIRs).  While not conclusive, both sets of data are 
consistent with models where the CIRs are either a source of X-rays or 
modulate them.
\end{abstract}

\begin{keywords}
stars:  individual ({HD\,24912}) ---
          stars:  early-type ---
          stars:  winds, outflows ---
	  X-rays: stars 
\end{keywords}

\section{Introduction}
X-ray emission is ubiquitous in the O stars and taken as an indication of 
dynamic instabilities in their winds \citep{lw80, l82, ow88, fel97}.
These 1-D hydrodynamical models predict a plasma with temperature $\sim 1$
--10\,MK, which is permeated with cool wind clumps.  The models also 
predict very strong stochastic X-ray variability, on time scales of hours.  
However, it has been clear since early X-ray observations, that stochastic 
variability on such short time scales is very small, less than about 1\%.  
To explain this, \citet{cas83} suggested that the winds contain thousands 
of shocks, and \citet{fel97} speculated that models with full 2-D 
hydrodynamics would reduce the predicted level of stochastic X-ray 
variability.  In the most extensive analysis to date, \citet{naz13} 
examined high quality {\em XMM-Newton} observations of the early O 
supergiant $\zeta$\,Pup.  They used an {\it ad hoc} 2-D wind model and 
found that the lack of stochastic X-ray variability on short time scales 
required a highly fragmented wind with a huge number of small clumps.  In 
this picture, a stellar wind consists of a large population of cool clumps, 
which contain the bulk of the stellar wind matter seen at UV, optical, IR 
and radio wavelengths, and a tenuous hot interclump medium responsible for 
the X-rays.  Further evidence for small scale clumping has come from 
the analysis of optical and UV wind lines.  \citet{hil91} found that it was 
necessary to introduce clumping to explain the shapes of electron scattering 
emission line wings in Wolf-Rayet (WR) stars.  Given the 1-dimensional 
nature of his model, this would imply the presence of either concentric 
shells or random structures whose size and separation are much smaller than 
the Sobolev length, so that the angle integrations are meaningful.
Stochastic variable features in the He\,{\sc ii} $\lambda 4686$\,\AA\ 
emission line in $\zeta$\,Pup were found by \citet{e98}, and explained as 
excess emission from the wind clumps.  \citet{mar05} investigated H$\alpha$ 
line-profile variability in a large sample of O-type supergiants, and 
concluded that the observed variability can be explained by a wind model 
consisting of coherent or broken shells.  \citet{lm08} presented direct 
spectroscopic evidence of clumping in O and WR star winds.

Besides this small scale clumping, a different, perhaps related, aspect of
O star winds is that all of them which have been observed over time scales
of a day or more demonstrate temporally coherent UV wind line variability
\citep[e.g.,][]{mega, k96, k99}.  This phenomenon suggests the presence of
large structures in their winds which {\bf may} originate from large regions 
on the surface of the star.  These results led \citet{co96} to model 
wind variability by large spiral structures known as Co-rotating Interaction 
Regions \citep[CIRs,][]{m86}.  These structures originate from large regions 
of enhanced wind flux on the surface of the star, although the exact cause  
of the enhanced wind remains unexplained.  \citet{ham01} showed that the 
observed UV line variability cannot be simply explained as a consequence of 
rotation in the framework of the CIR model, and a more complex interplay 
between rotation and radial velocity is taking place.  Further, \citet{pm10} 
analyzed the doublet ratios of wind lines of a large number of B 
supergiants.  Their results demonstrated that the spectral signature of 
large, optically thick wind structures which cover only a portion of the 
line of sight to the stellar disk (most likely CIRs) is common in the B 
supergiants.

The interplay between CIRs and small scale clumping is largely unexplored.  
A crude analysis by \citet{o99} suggested an intricate interaction between 
the two.  \citet{lb08} considered 3-D hydrodynamic models of CIRs.  While 
their models could reproduce the detailed time evolution of UV spectral 
features in a B-type supergiant, they expressed concern that too much small 
scale clumping could destroy the CIRs.  Whatever the mechanism of X-ray 
production, the presence of large scale structures in stellar winds 
should leave a footprint on the X-ray emission.  In particular, X-ray 
variability on a time scale compatible with the stellar rotation period 
should be present, and this X-ray variability should correlate with UV wind 
line variability.  However, there are considerable observational obstacles 
to establishing a link between X-ray and UV wind line variability for 
single, normal O stars.  First, the CIRs are thought to be confined to the 
equatorial plane.  Consequently, variability may not be seen unless our 
line of sight to the star is relatively close to equator-on.  Second, 
the stellar rotation period for most O stars is several days.  This is 
considerably longer than the typical X-ray observation, so variability 
could easily be missed.  Third, there are only a few stars which have been 
observed in the UV over long enough intervals for good wind line periods to 
be identified, and such a period is needed to claim a definite connection 
between X-ray and UV wind line variability.  Fourth, binary stars with 
colliding stellar winds must be excluded from any investigation of the 
connection between X-rays and large scale wind structure.

Despite these hurdles, observational evidence supporting a link between 
CIRs and X-ray emission from stellar winds is mounting.  Recently, X-ray 
variability of $\sim 20$\% and on the time scale of recurrent DACs was 
detected in the WR star EZ CMa, and explained in the framework of the CIR 
model \citep{o12, i13}.  More relevant to the current paper, is the work on 
O stars by \citet{berg96}, \citet{o01} and \citet{naz13}.  \citet{berg96} 
found a marginal detection of a periodic X-ray variability in $\zeta$~Pup, 
which was not confirmed by \citet{o01} or \citet{kah01} and not in 
agreement with DAC period determined by \citet{how95}.  \citet{naz13} 
analyzed a series of X-ray observations of $\zeta$~Pup which span 10\,years.  
They detected a slow modulation of the X-ray flux with a relative amplitude 
up to 15\%\ over 16 hours in the 0.3-4.0~keV band.  They propose that these 
modulations can be attributed to CIRs.  The most compelling evidence to 
date was given by the \citet{o01} analysis of the rapidly rotating O dwarf 
$\zeta$\,Oph.  It is the only single O star that has been observed 
continuously in X-rays over a full rotational period.  These observations 
showed that the X-ray variability occurred on a time scale similar to the 
UV wind line variability, but they were not long enough to show that
the pattern repeated on the rotation time scale.  Unfortunately, there were
no simultaneous observations of the wind activity that could be used to 
determine the phase relation between the X-ray and wind activity.  Such a 
relationship could provide valuable clues about the geometric relationship 
between the X-ray emitting plasma and the CIRs.

In this paper we present X-ray observations covering nearly half of a 
rotation period of the O7~III(n)((f)) star ${\xi}$~Per, supplemented by 
optical \ha\ observations which bracket the X-ray observations.   
${\xi}$~Per is an ideal candidate for attempting to make a connection 
between UV wind line and X-ray variability.  First of all, $\xi$~Per 
appears to be a perfectly normal O7 giant \citep[see,][for descriptions of 
its optical and UV spectra, respectively]{w73,w85}.  Its only 
distinguishing feature is that it has a somewhat high (but not abnormal) 
rotational velocity of $v\sin i = 204$ km~s$^{-1}$ \citep{p96}.  Because of 
its moderate $v \sin i$, its expected rotation period is short enough to be 
captured by a time series of manageable duration.  This was done in a 
detailed study of its UV and optical line variability by \citet{deJong}.  
They demonstrated the presence of a well characterized, distinctive 
2.086~day period in the UV wind line variability.  This turns out to be 
roughly half of the expected rotation period if the star is viewed nearly 
equator-on (see their Figure~4).  Further, they were able to model the 
variability by two sets of two armed CIRs (see, their Figure~17), with one 
set of arms dominating the variability.  They were also able to establish 
a relationship between the appearance of the discrete absorption components 
(DACs) associated with the CIRs and variations in the blue wing of \ha, 
which enable us to use \ha\ variability as a surrogate for DAC activity.  
These properties of $\xi$~Per, together with its relative brightness at 
X-ray wavelengths, make it an ideal candidate to determine whether the CIRs 
that are thought to modulate the UV wind lines in this star (and possibly 
all luminous OB stars with massive winds) might modulate its X-ray flux as 
well.  To pursue this conjecture, we obtained a 162~ks (0.45 rotation 
periods and 0.90 periods of the strong CIR activity) \chandra\/ observation 
of $\xi$~Per.  

The X-ray and optical observations are described in \S\ref{observations}, 
anlyzed for variability in \S\ref{analysis} and discussed in 
\S\ref{discussion}. 

\section{The Observations and Data Reduction}\label{observations}

\subsection{X-ray Data}
Although a continuous observation of 2.1~d is required to be certain that 
the full range of variability is observed, as one arm of the CIR pattern 
traverses our line of sight, the maximum visibility of $\xi$~Per to 
\chandra\/ was $\sim 162$~ks (1.9~d), or roughly 90\% of the UV wind 
line modulation period.  The observations were obtained with \chandra's 
transmission High Energy Transmission Grating Spectrometer (HETGS) and 
began on 2004 March 22 (${\rm MJD} = 53086.0904$).  

The high energy transmission grating, HETG, consists of two sets of 
gratings; the Medium Energy Grating (MEG) and the High Energy Grating (HEG) 
\citep{chanobs}.  The events were recorded on the ACIS-S chips and consist 
of a $\pm$ first order dispersed spectrum for both the MEG and HEG and a 
directly transmitted, zeroth order, image.  Some spectral information can 
be extracted from the zeroth order image through the energy determined for 
the individual events.  All of the data were processed with version 3.3 of 
the \chandra\/ Interactive Analysis of Observations (CIAO) software 
package.  The newly reduced data can be found in the ''The \chandra\/ 
Grating-Data Archive and Catalog (TGCat)'' \citep{ho11}.  See the CIAO 
documents for further 
information\footnote{http://cxc.harvard.edu/ciao/intro/}.

\begin{figure}
\begin{center}
\epsfig{figure=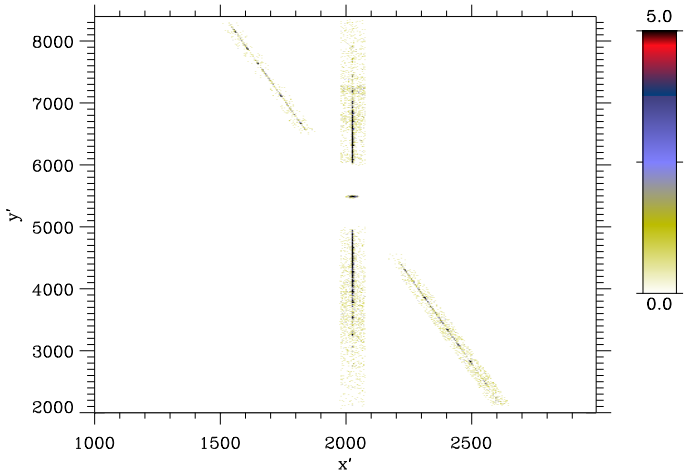,width=3.5in}
\end{center}
\caption{Partial image of the \chandra\ data of $\xi$~Per in a coordinate 
frame rotated so that the MEG spectrum is a vertical stripe.  This image 
contains only events filtered as described in the text.  The spot near the 
center of the image is the zeroth order image and the faint diagonal stripe 
is the HEG spectrum.  
\label{fig:image}
}
\end{figure}
\begin{figure}
\begin{center}
\epsfig{figure=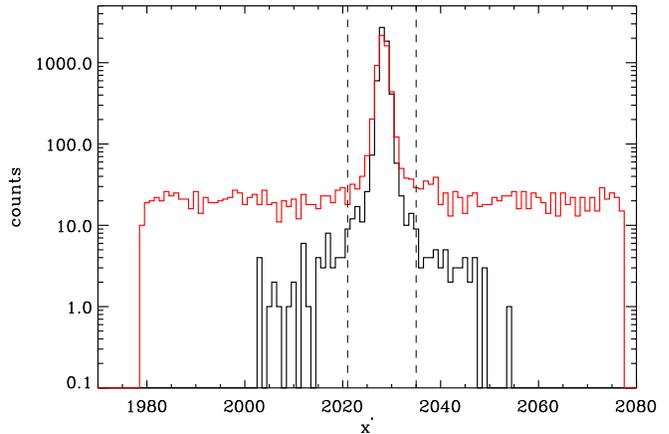,width=3.5in}
\end{center}
\caption{Cross dispersion profiles for the $M = \pm1$ MEG (red) spectra and 
$M=0$ (black) image.  Each profile is the total number of counts for the 
order collapsed along the $y^\prime$ axis shown in Figure~\ref{fig:image}.  
The filtering of the data is described in the text. \label{fig:profs}
}
\end{figure}

Pile-up (the arrival of two photons near the same location between CCD 
reads) was not a major issue for the observations.  The strongest lines in 
the MEG spectra only had count rates of $\lesssim 1.0\times 10^{-3}$~counts 
s$^{-1}$ and the count rate of the zeroth order image was $\lesssim 
0.04$~counts~s$^{-1}$.  In the first case, pile-up is completely 
negligible, and in the latter it never exceeds 5\%.  Further, pile up will 
only mute the amplitude of variability and not make a constant source 
appear variable. 

We now describe how we filtered the initial data to maximize the 
contribution from the source and minimize background effects.  We 
utilize the \chandra\ primary time tag photon list for our analysis.  
Specifically, we use the following tags attached to each event: $x$ and 
$y$, the sky coordinates; $energy$, the event energy estimated from the 
number of CCD counts it created; $grade$, the shape descriptor of the CCD 
count distribution of the event; $time$, the time at which the event was 
read; $tg\_m$, the likely grating order, $M$, associated with an event; 
$tg\_lam$, the wavelength for the $M = \pm1$ events based on their location 
relative to the dispersion, and; $tg\_part$, the probable association of a 
photon with either the HEG or MEG spectrum of the zeroth order image based 
on its location.  See, the CIAO web site for further information.  

To simplify matters, we begin by ``rotating'' the $x$ and $y$ sky 
coordinates by $\theta = -29.31^\circ$, to produce a primed coordinate
system in which the first order MEG spectra are aligned with the 
$y^\prime$-axis.  Figure~\ref{fig:image} shows a portion of the detector in 
the rotated frame.  The events in this image have been filtered as follows: 
$|M| \leq 2$, ASCA grades of 0, 2, 3, 4, 5 or 6 (as done in normal 
processing), and $0.4769 \leq energy \leq 2.0664$~keV, which corresponds to 
the wavelength range of interest, $6 \leq \lambda \leq 26$ \AA.  The $M = 
0$ image appears as a spot near the center of the figure, and the MEG $M = 
\pm 1$ spectra are the vertical strips which lie on either side of it.  The 
faint diagonal stripes are the HEG spectra, which are not considered 
further, since they contain very few counts.  

We note two important properties of Figure~\ref{fig:image}.  First, the 
zeroth order image is localized and resides on a single CCD.  Second, in 
spite of filtering, there are considerable extraneous events in the 
cross-dispersion direction, well away from the center of the MEG $M = \pm 
1$ spectra.  To emphasize this, we filtered the events list further to 
produce cross-dispersion profiles.  Figure~\ref{fig:profs} shows the cross 
dispersion profiles for the MEG $M = \pm 1$ events collapsed along the 
$y^\prime$ axis, compared to all of the $M = 0$ events, similarly 
collapsed.  The first order events were restricted further to include only 
counts with $tg\_part = 2$ (which eliminates HEG data) and wavelength tags 
in the range $6 \leq \lambda \leq 26$\AA.  This last restriction differs 
from the energy filtering since it also includes location on the detector.  

Because the events on the wings of the first order cross-dispersion profile 
do not appear in the zeroth order image, they are not from the sky.  
This is further verified by examining their spectral signature, which turns 
out to be featureless -- free of emission lines.  Therefore, these counts 
will not carry the time dependence of the source and we should eliminate as 
many as possible.  To do this, we restrict the first order events to those 
which lie in the range $2021 \leq x^\prime \leq 2035$.  There are 5782 
counts in the refined region, compared to 7580 when all $x^\prime$ events 
are included -- a reduction of 31\%.  Assuming that the source counts are 
all contained in the restricted area and that the background counts are 
uniformly distributed over the region $1979 \leq x^\prime \leq 2077$ (see 
Figure~\ref{fig:profs}) we can estimate the total number of background and 
source counts.  Let $N_s$ and $N_b$ be the total number of source and 
background counts over the region from $1979 \leq x^\prime \leq 2077$.  
Now let $N_1 = 7580$ be the observed counts over the larger region and 
$N_2 = 5782$ be the observed counts over the restricted region.  Then, 
$N_1 = N_s + N_b$ and $N_2 = N_s + N_b \times(2035-2021)/(2077-1979)$.  
Solving these two equations gives $N_s = 5482$ and $N_b = 2098$.  Thus, 
28\% of $N_1$ are background counts compared to only 5.2\% of $N_2$.  
Consequently, although some background counts remain in $N_2$, their effect 
should not be too great.  In contrast, the $M = 0$ image has 5915 counts 
and appears to contain far fewer background counts.  Therefore, it was not 
filtered any further.  

\begin{figure}
\begin{center}
\epsfig{figure=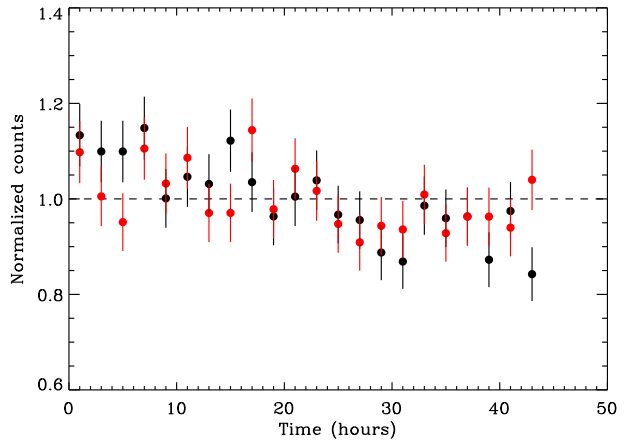,width=3.5in}
\end{center}
\caption{Total counts in the $M = \pm 1$ MEG spectra (red) and the $M=0$ 
(black) image resampled into 2~hour bins and normalized by their mean 
counts.  The error bars are based on the total number of counts contained 
in each time bin.
\label{fig:compare}
}
\end{figure}

\begin{figure}
\begin{center}
\epsfig{figure=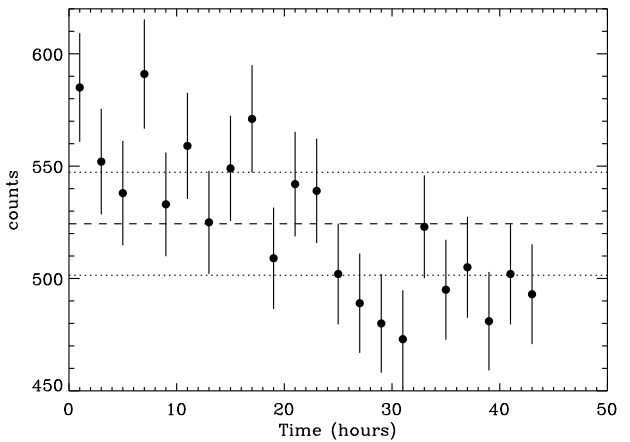,width=3.5in}
\end{center}
\caption{Total counts in the combined $M = \pm1$ and $M = 0$ MEG spectra 
resampled into 2~hour bins.
\label{fig:combined}
}
\end{figure}

The filtered events from the zeroth order image and first order spectra 
were resampled into 2 hour time bins for better statistics.  
Figure~\ref{fig:compare} compares the two time series.  In this figure, 
each series was normalized by its mean value for comparison purposes and 
the error bars are based on Poisson statistics from the number of counts in 
each bin.  It is clear that both series display a general decreasing trend 
over the extent of the observation.  Therefore, the two were added to 
arrive at the summed series shown in Figure~\ref{fig:combined}, where each 
bin contains several hundred counts, making it reasonable to approximate 
the intrinsic Poisson statistics by Gaussian statistics. 

\subsection{Optical Data}

While the period of DAC variability in $\xi$~Per is reasonably well defined, 
it probably originates from surface features, and there is no reason to 
expect long term coherence.  Consequently, to determine the phase relation 
between the X-ray and DAC variability at the time of the \chandra\ time 
series, we need observations which predict DAC occurrence and are obtained 
close to the time of the X-ray observations.  To do this, we use the 
\citet{deJong} result that variable absorption of the blue wing of \ha\ is 
essentially in phase with the DACs (see their Table~5).  Consequently, \ha\ 
observations were obtained within $\sim \pm 7$ days of the \chandra\ data, 
in order to determine the relative phases of the X-rays and the DACs.  
Spectra were obtained at three North American sites: the David Dunlap 
Observatory (DDO); the Ritter Observatory (Ritter); and the Dominion 
Astrophysical Observatory (DAO).

Our optical observations bracket and overlap the X-ray observations.  They
begin 6.6 days prior to the \chandra\ observations and conclude 7.7 days
after.  They span a total of 16.1 days, roughly centered on the \chandra\
series.  Since \citet{deJong} have demonstrated that the CIR related
spectral features remain well phased over the 9.4 day interval of their
{\it IUE} time series, it seems reasonable to assume that they remain in
phase over the time interval of our optical data, which is less than twice
as long.

\noindent \underline{{\bf DDO Observations:}} 
One high-quality spectrum was obtained with the Cassegrain spectrograph of
the DDO 1.88m telescope.  The observation was made through a 242~$\mu$m 
slit with the 1800 lines/mm grating in first order. The detector was a 
Jobin-Yvon thinned, back-illuminated CCD with 2000$\times$800 pixels.  
This configuration provided a linear reciprocal dispersion of 0.153~{\AA} 
per 15~$\mu$m pixel.  Bias frames, tungsten lamp flat field observations 
and Fe-Ar lamp wavelength observations were obtained immediately before and 
after the stellar observation.

\noindent \underline{{\bf Ritter Observations:}} 
Three spectra were obtained with the Ritter 1.06~m telescope. These were 
obtained with the bench-mounted {\'e}chelle spectrograph, coupled to the 
Cassegrain focus of the telescope by a 200~$\mu$m wide optical fiber.  The 
detector was a front-illuminated EEV CCD with 1200$\times$800 pixels, each 
of which is 22.5~$\mu$m square.  At {\ha} (order 34), the spectrograph has 
resolving power of $2.6 \times 10^4$, and the entrance slit projects to 
$\sim$4.3 pixels.  The observed positions of telluric water vapour lines 
were first used with the wavelengths given by \citet{hetal2000} to 
correct nonlinearities in the wavelength scale determined from the 
comparison arc.  After telluric correction, the Ritter spectra were 
binned by a factor of 2 to improve the S/N.

\noindent \underline{{\bf DAO Observations:}} 
The DAO 1.22m McKellar telescope and coud{\'e} spectrograph were used in 
``robotic'' mode to obtain seventeen spectra on 4 nights. The ``9681M'' 
configuration of the spectrograph was used with the SITe-4 CCD detector, 
which is a thinned, back-illuminated array of 4096 $\times$ 2048 pixels, 
each of which is 15 $\mu$m square.  This configuration results in a 
spectrum with linear reciprocal dispersion of 0.073~{\AA}/pixel.  
Additional observations of a tungsten continuum source and Th-Ar emission 
spectrum were obtained to provide flat field and wavelength calibrations, 
respectively.

Standard {\it IRAF}\footnote{{\it IRAF} is distributed by the National 
  Optical Astronomy Observatories, which are operated by the Association of 
  Universities for Research in Astronomy, Inc., under cooperative agreement 
  with the National Science Foundation.} 
routines were used to process the CCD frames obtained at all three
observatories.  Telluric line contamination was removed by using templates 
derived from contemporaneous spectra of the rapidly rotating B7~V $\alpha$ 
Leo.

The late March date of the \chandra\ observations had two consquences for 
the ground-based campaign.  First, {\xper} could only be observed for
$\lesssim$2~hours at the beginning of each night.  Consequently, a 
significant phase range of the 2.086-day period could not be covered during 
a single night, and nearly the same phase was observed every second night.  
Second, the weather is not traditionally very good at any of the 
observatories in March.  As a result, the phase coverage is fairly sparse.  

\begin{figure}
\epsfig{figure=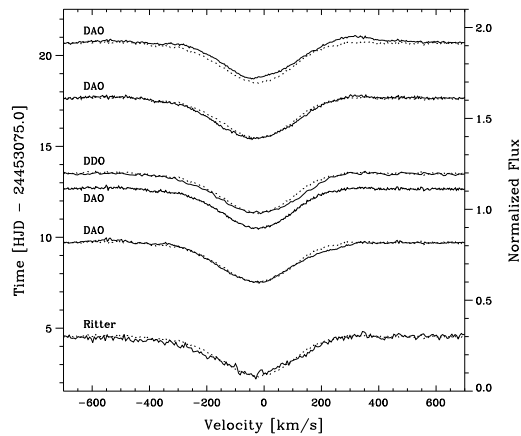, width=3.5in}
\caption{Nightly mean {\ha} spectra obtained contemporaneously with the 
{\it Chandra} X-ray observations.  The abscissa denotes velocity in the 
frame of the star, for which a systemic velocity of {59.3~\kms} has been 
adopted.  Spectra are displaced vertically in proportion to the time 
difference between them, as indicated by the left-hand scale.  The 
right-hand scale indicates normalized flux.  To highlight changes, the 
mean DAO spectrum from 2004-03-23 is overplotted as a dotted line.  
Two Ritter spectra have not been 
plotted in order to avoid confusion with the other observations obtained 
on the same nights. \label{hastack}
}
\end{figure}

Since the phase coverage on any given night was extremely limited, mean 
spectra were computed whenever possible in order to improve the 
signal-to-noise (S/N) ratio.  The nightly mean spectra are illustrated in
Figure~\ref{hastack}.  We define the ``blue wing'' equivalent width of 
{\ha}, $W_B$(\ha), as the integrated mean spectra between $(-500, 
-200)$~{\kms} of line center (measured in the stellar reference frame). The 
prescription of \citet{ve06} was used to estimate uncertainties in 
$W_B$(\ha).

Table~\ref{journal} summarizes the optical data.  Successive columns record 
the heliocentric Julian Data (HJD) corresponding to the mid-point of the 
mean spectrum; the observatory; the number of spectra averaged; the 
exposure time per spectrum in seconds; the S/N per pixel in the continuum 
of the mean spectrum; and $W_B$(\ha) in {\AA}.

\begin{table}
 \centering
 \begin{minipage}{140mm}
  \caption{Journal of Ground-Based Observations\label{journal}}
  \begin{tabular}{lcrrcc} \hline 
HJD\,\footnote{HJD $-$ 2,400,000.0 at the mid-point of the mean spectrum.} 
& Obs. & N\,\footnote{Number of spectra.} & t$_{\rm exp}$ [s] &
S/N\,\footnote{S/N per pixel in the continuum of the mean spectrum.} &    
$W_B$(\ha) [\AA]\,\footnote{Measured between $(-500, -200)$~{\kms}.} \\ 
\hline
53079.5386 & Ritter & 1 & 1987 & 112 &   0.25 $\pm$ 0.08  \\
53084.7081 & DAO    & 4 & 1200 & 397 &   0.13 $\pm$ 0.02  \\
53087.5534 & Ritter & 1 & 3600 & 187 &   0.17 $\pm$ 0.05  \\
53087.6730 & DAO    & 3 & 1200 & 264 &   0.15 $\pm$ 0.04  \\ 
53088.5196 & DDO    & 1 & ~212 & 211 &   0.20 $\pm$ 0.04  \\
53088.5383 & Ritter & 1 & 3600 & ~89 &   0.18 $\pm$ 0.10  \\ 
53092.6395 & DAO    & 1 & 1200 & 238 &   0.17 $\pm$ 0.04  \\
53095.6826 & DAO    & 7 & 1200 & 385 &   0.06 $\pm$ 0.02  \\  
\hline
\end{tabular}
\end{minipage}
\end{table}

\section{Analysis}\label{analysis}

The mean HETGs spectrum of $\xi$~Per was already presented by 
\citet{ofh2006} and \citet{w09}. The X-ray spectrum is similar to other 
single O-type stars whose X-ray line profiles can be described in the 
context of a clumped wind model with the X-ray emission originating 
relatively close to the stellar photosphere. Therefore, this work 
concentrates on the time variability.

If we simply take the standard deviation of the time series shown in 
Figure~\ref{fig:combined}, we find that the probability that the scatter 
about the mean exceeds the observed value is 15\%.  Based on this, one 
might conclude that the evidence for variability is not very strong.  
However, this statistic does not consider the distinct temporal ordering of 
the deviations.  The object of this section is to analyze the observed 
trend.  We first assess the time variability by subjecting the un-binned 
data to the Kolmogorov-Smirnov test.  Next, we examine the form of the 
autocorrelation function \citep[e.g., ][]{chat04} of the binned data.  We 
then apply three different models for the time dependence.  Next, we examine 
a hardness ratio of the data for variability.  Finally, we examine the 
variability of the \ha\ spectra. 

Unlike a simple $\chi^2$ statistic, the Kolmogorov-Smirnov test accounts 
for the distribution of the variance in time.  We applied the 
Kolmogorov-Smirnov test to compare the unbinned zeroth and first order 
events, subject to the final screening outlined in the previous section, 
to a uniform, non-variable, sequence.  This test resulted in probabilities 
of 0.9984 and 0.9233, respectively, that the data were time variable.  The 
difference probably results because the first order spectra contain more 
background contamination.  The same test was applied to the combination of 
the two data sets and gave a probability of 0.9997 that it is variable.  
These results leave little doubt about the reality of the temporal 
variability.

\begin{figure}
\begin{center}
\epsfig{figure=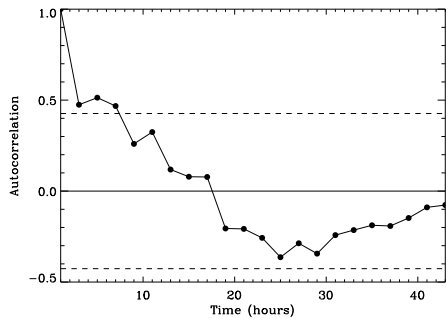,width=2.5in}
\end{center}
\caption{Autocorrelation function of the time series shown in 
Figure~\ref{fig:combined}.
\label{fig:auto}
}
\end{figure}

To examine the time dependence more closely, we present the autocorrelation 
function of the binned time series in Figure~\ref{fig:auto}.  The dashed 
lines are $\pm 1/\sqrt{22}$, which are the $\pm 2\sigma$ limits for the 
expected values of the elements of a random sequence with 22 samples 
\citep{chat04}.  The fact that the first 3 autocorrelation coefficients 
exceed this value is also strong evidence that the series is time variable.  
Further, the shape of the autocorrelation function bares no resemblance to 
one expected from a random sequence.  Instead, it is more indicative of a 
portion of a periodic signal with its broad, negative minimum near 25 
hours (roughly half of the period), in agreement with expectations.  

We next modeled the binned series shown in Figure~\ref{fig:combined} with 
three heuristic models: 
\begin{enumerate} 
\item A linear model, which measures the overall trend in the data
\begin{equation}
{\rm counts} = a_1 t+a_2  \;.
\end{equation}

\item A trigonometric model, to search for evidence of repeatability
\begin{equation}
{\rm counts} = a_1 \cos \left(\frac{t -a_2}{2\pi P} \right) +a_3  \;, 
\label{eq:trig}
\end{equation}
where $P$ is fixed at 2.086d \citep{deJong}. 

\item An exponential decay model, to characterize the recovery from an 
episodic event
\begin{equation}
{\rm counts} = a_1 e^{-t/a_2} +a_3  \;.
\end{equation}

\end{enumerate}

\begin{figure}
\epsfig{figure=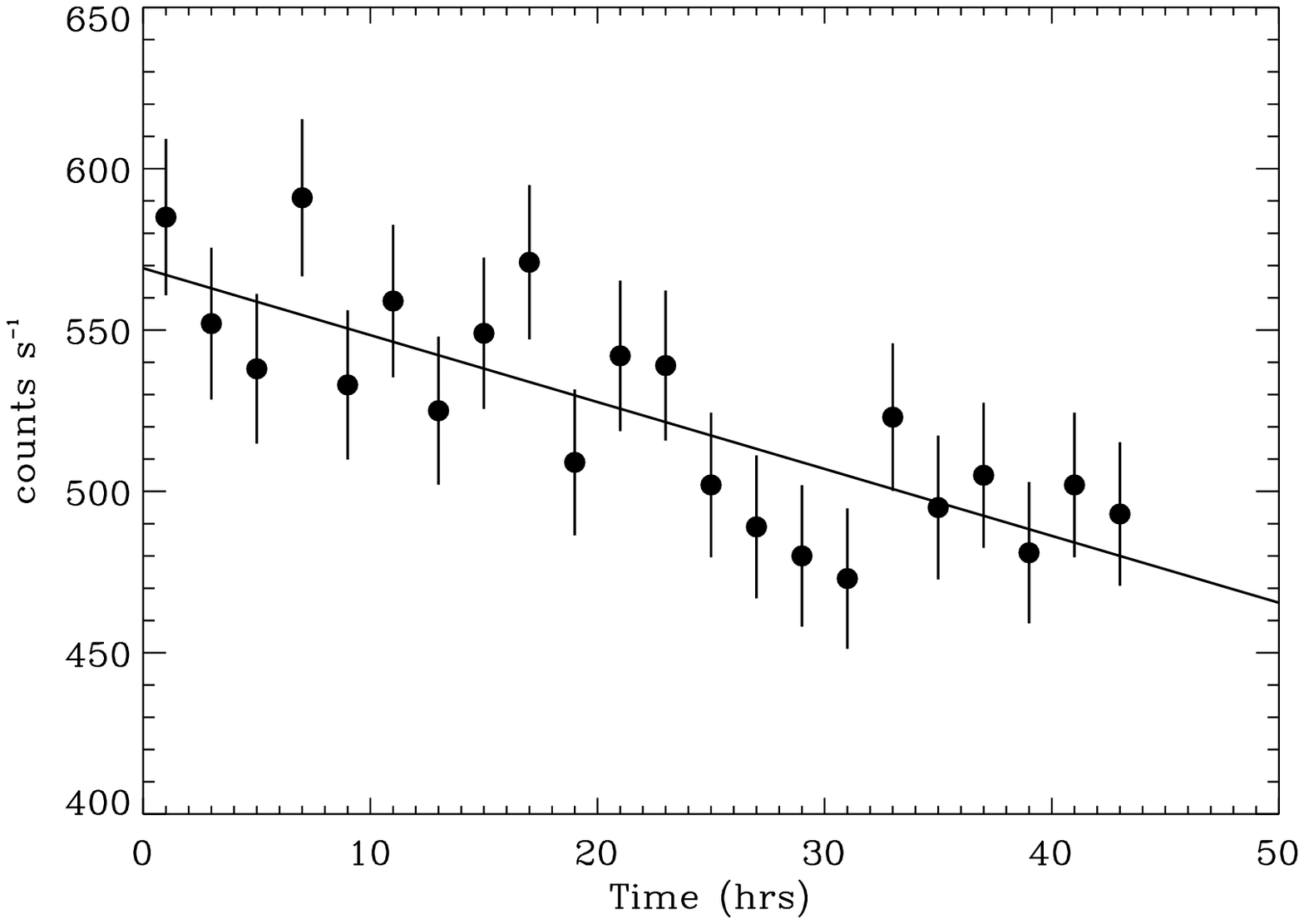,width=3.0in}\hfill
\epsfig{figure=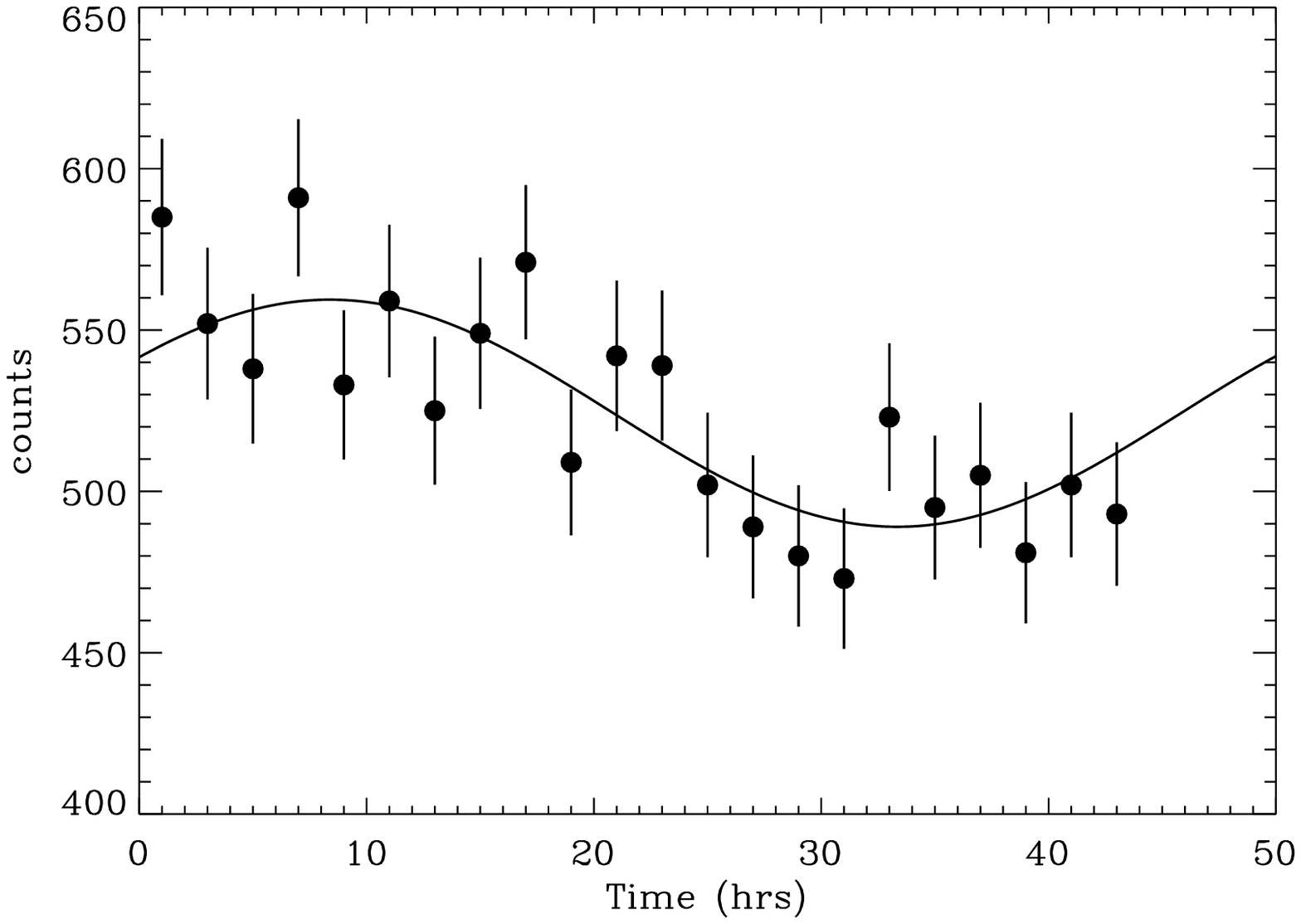,width=3.0in}
\caption{Fits of the total counts shown in Figure~\ref{fig:combined} to a 
linear function (top) and a trigonometric function (bottom).  The fit 
parameters are given in Table~\ref{tab:fit_params}.
\label{fig:fits}
}
\end{figure}

The results of this exercise are shown in Figure~\ref{fig:fits} and are 
summarized in Table~\ref{tab:fit_params}.  The top panel shows the linear 
fit and the bottom panel the cosine fit, replicated to demonstrate its 
consistency with a periodic function.  We do not show the exponential fit 
since it is indistinguishable from the linear fit.  
Table~\ref{tab:fit_params} lists the fit parameters.  There are two rows 
for each fit.  The first row gives the name of the fitting function, the 
free parameters and the probability that the observed reduced $\chi^2$ 
exceeds the value expected if the assumed functional form were exact.  The 
second row gives the mono-variate uncertainties of the fit parameters.  

\begin{center}
\begin{table}
\caption{Fit Parameters}
\begin{tabular}{lrrrc} \hline
Function  & \multicolumn{1}{c}{$a_1$} & \multicolumn{1}{c}{$a_2$} &  
 \multicolumn{1}{c}{$a_3$} &  \multicolumn{1}{c}{Probability} \\ \hline
linear      & $ 569.2$    & $-2.07$    & \multicolumn{1}{c}{--}& 0.59 \\ 
            & $\pm 10.0$  & $\pm 0.39$ & \multicolumn{1}{c}{--}  &  \\ 
cosine      & $35.8$      & $7.9$      & $520.8$     & 0.62 \\
            & $\pm 6.5$   & $\pm 1.7$  & $\pm 5.0$   &      \\ 
exponential & $573.0$     & $251.0$    & $-2.0$      & 0.53 \\ 
            & $\pm 3281$  & $\pm 1585$ & $\pm 3387$  &      \\
\ha\ cosine & $0.082$     & $25.1$     & $0.150$     & 0.45 \\
            & $\pm 0.022$ & $\pm 1.6$  & $\pm 0.014$ &      \\ \hline
\end{tabular}
\label{tab:fit_params}
\end{table}
\end{center}

\begin{figure}
\begin{center}
\epsfig{figure=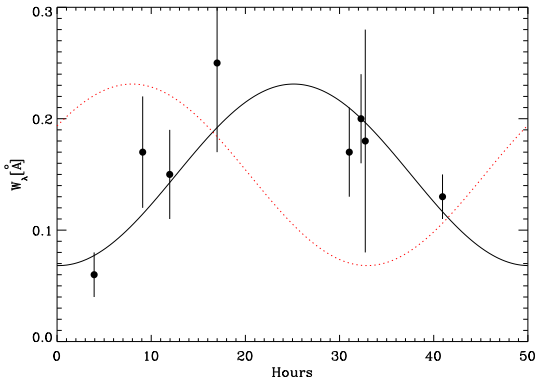,width=3.0in}
\end{center}
\caption{The strength of \ha\ blue-wing absorption fit by the cosine 
function given by eq.~\ref{eq:trig}.  The fit parameters are listed in 
Table~\ref{tab:fit_params}.  The red curve is the same fit, except with the 
X-ray value for the phase, $a_2$.  The optical data have been wrapped on the 
2.086-day period and phased to the beginning of the X-ray time series.
\label{fig:HaX}
      }
\end{figure}

The reduced $\chi^2$ statistic indicates that the cosine function provides 
the best fit, but that it is not significantly better than the linear or 
exponential functions.  The coefficient errors are extremely large for the 
exponential because they are strongly correlated, i.e., the coefficients 
can be changed together by large, but related, amounts and still produce a 
good fit.  

Finally, we examined the time dependence of the hardness ratio.  This was 
done by dividing the combined MEG $M = \pm 1$ and $M = 0$ spectrum into two 
bands: a ``soft'' band (events with $0.477 \leq {\rm h}\nu < 0.954$~keV), 
and; a ``hard'' band (events with $0.954 \leq {\rm h}\nu < 2.066$~keV).  
This division gives comparable counts in each band.  We then examined the 
time dependence of their ratio in 2 hour bins for variability.  The errors 
for the ratios were calculated using standard propagation of errors.  In 
this case, no significant evidence for variability was found.  For example, 
dividing the high energy bin by the lower energy bin gives $0.790 \pm 
0.026$ over the first half of the observation and $0.811 \pm 0.027$ over 
the second half.  Because the bulk wind opacity scales as $\sim 
\lambda^{-3}$, if an overall variability $\sim 15$\% is due to variable 
absorption (as might be caused by periodic occultation by the spiral CIR 
pattern), one might expect an even larger variation at shorter wavelengths.  
However, this is not observed.  One possible explanation for the 
absence of wavelength dependence of the variability is optically thick 
structures which occult different fractions of the source of the X-rays.  
If this is the case, the X-ray variability and its lack of 
wavelength dependence may provide powerful constraints on the geometry of 
CIRs and the source of the X-rays.  

We now turn to the \ha\ data.  Figure~\ref{fig:HaX} shows the $W_B$(\ha) 
data fit by the same cosine function used for the X-rays along with the 
cosine function phased to the X-rays.  The fit parameters are listed in 
Table~\ref{tab:fit_params}.  Although the data are sparse and the error 
bars large, formally, the fit is quite good.  The \ha\ and the X-rays are 
17.2 hours, or 124$^\circ$, out of phase.  This means that when the \ha\ 
absorption is strongest, the X-ray emission is nearly at its weakest.  

The major results of this section are that the X-ray spectrum of $\xi$~Per 
is variable at the level of $\sim 15$\% (determined from the 
cosine fit parameters), that the variability is consistent with modulation 
by the 2.086~day period associated with the CIRs, that no variation in the 
hardness of the spectrum could be detected, suggesting that the variability 
is due to partial obscuration by optically thick structures and that the 
magnitude of $W_B$(\ha) is distinctly out of phase with the intensity of 
the X-rays.  

\section{Discussion}\label{discussion}

There are basically two ways to generate X-ray variability.  One is {\it 
via} an impulsive event, and the other is by occulting the source with 
absorbing gas or the stellar disk, as in the case of $\zeta$~Oph 
\citep{o01}.  However, if the variability were due to an impulsive, 
flare-like event, it must have occurred long before the observations (since 
the form of the variability is nearly linear, indicative of an exponential 
tail).  This implies that the event would have been quite strong.  
Another possibility would be a flare on an unseen pre-main sequence 
companion.  However, the X-ray luminosity of $\xi$~Per is $\simeq 1.2\times 
10^{32}$ erg sec$^{-1}$ \citep{ofh2006}, while the X-ray luminosities of PMS 
stars are typically a factor of 10  smaller \citep[see,][]{pf05}.  Further, 
the observed spectral change is grey and soft, very uncharacteristic for 
flare spectra \citep[e.g.][]{gn09}.  While not impossible, either scenario 
seems unlikely, especially since the form of the variability and its 
autocorrelation function are consistent with cyclical behavior.  As a 
result, we favor the explanation that the X-rays vary because of 
obscuration by intervening material.

In addition to $\xi$~Per, X-ray variability has been observed in two 
other single, non-magnetic O stars: $\zeta$~Pup \citep[O4~If(n),]
[]{naz13}, and $\zeta$~Oph \citep[O9.5 Vnn,][]{o01}.  Both vary on time 
scales which may be related to their stellar rotation periods.  An analysis 
of several years of X-ray data for $\zeta$~Pup showed evidence for 
variability on a time scale similar to its DACs, but no single time series 
encompassed a rotation period.  The best evidence for the interaction 
between CIRs and X-rays to date is the $\zeta$~Oph data set.  It was 
observed for $\sim 1.2$ days and its X-ray flux was found to vary by about 
20\%.  Further, this variation appeared to repeat with a period of about 
0.77 days, roughly half of the stellar rotation period and similar to a 
period previously determined from its UV wind lines ($0.875 \pm 
0.167$~days) by \citet{h93}.  However, no information relating the phases 
of the X-rays and the DACs was available, so it was impossible to constrain 
the geometry of the CIRs and X-rays.    

In this paper, we have presented three observational results which help 
constrain the relation of the CIRs and X-rays.  These are: X-ray 
variability, consistent with the period observed in the DACs; a constant 
hardness ratio for the X-rays, suggesting that the source of the X-rays 
becomes partial obscured by optically thick structures along the line of 
sight, and; a phase lag of 124$^\circ$ between the maximum absorption on 
the blue wing of \ha\ and the maximum strength of the X-rays.  

To interpret our limited set of observations (covering roughly half of a
rotation period) in terms of the geometry of CIRs requires some constraints.  
We follow \cite{deJong} and adopt the following 3 assumptions:
\begin{enumerate}
\item {\bf The wind of $\xi$ Per contains two pairs of spiral arms, with one 
arm in each pair being much stronger than the other.  We concentrate on the 
two major arms, which are equally spaced.}

\item The wind structures follow streak lines whose shapes are defined by
the stellar rotational velocity, wind velocity law and terminal velocity.
We adopt a rotational velocity equal to the observed $v\sin i = 205$~\kms,
a velocity law of the form $v = v_\infty [1 -a/(r/R_\star)]$, where
$v_\infty = 2420$~km~s$^{-1}$ and $1 -a = v(r = 1) = 0.01 v_\infty$.

\item The \ha\ emission originates near the base of the wind and is 
associated with the spiral structures.  Since increased emission reduces 
$W_B$(\ha), the \ha\ emission and X-ray variability are nearly in phase. 
\end{enumerate}
In addition, we note that the grey nature of the variability suggests that 
the X-ray variations are due to occultation of the X-ray source by either 
the stellar disk or wind structures that are optically very thick to X-rays.

Thus, we seek configurations where the X-ray emission and \ha\ emission are
roughly in phase.  However, even with these restrictions, there is still
considerable latitude in how the geometry can be arranged.  We consider two
possibilities, but others may be possible.

\begin{figure}
\vspace{-0.5in}
\vspace{-0.1in}\epsfig{figure=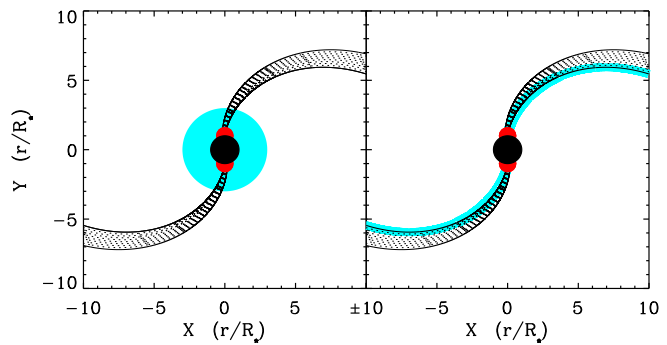,width=3.7in}
\vspace{-0.3in}\caption{
Cartoons of two different configurations of the \ha\ and X-ray emitting 
regions in a wind containing two spiral structures in the shape of streak 
lines.  In each case, the \ha\ emitting region is depicted by two red dots, 
and the X-ray emitting region by an aqua region.  The coordinates are in 
units of $r/R_\star$.\label{fig:cartoons}}
\end{figure}

The right panel of Figure~\ref{fig:cartoons} shows the first configuration.
It places the X-ray emission along the interacting edge of the spirals.
This might be the case if some or most of the X-rays originate at the fast
wind--slow wind interface, which creates the CIRs.  Further, because the
velocity differential between the fast and slow winds (and, presumably the
potential to produce X-rays) is expected to drop with distance from the
star, one expects most of the X-rays to originate within $\sim 5 R_\star$.
Thus, viewing the Figure from the right, we see that the \ha\ and
X-ray emitting regions are at or near maximum.  Then, as the configuration
rotates counter-clockwise, our view would be from below.  In this case, one
lobe of the \ha\ emission and much of the X-ray emission near the star are
occulted by the stellar disk, producing a minimum in both.  Finally, as the
system rotates further and our view is from the left, and we have returned 
to a maximum once again.

For this configuration to explain the observations, the arms must represent
a relatively low density contrast in the wind since we do not observe
spectral changes in the X-ray spectrum indicative of absorption.  This is
in accordance with typical CIR models, where UV DAC formation is attributed
mostly to a velocity plateau.  However, in contrast to the classic line
driven instability (LDI) model (which predicts that the X-ray source should
be distributed relatively uniformly throughout the wind) the observed
variability implies that the CIRs must account for a large fraction of the
X-rays.

The left panel of Figure~\ref{fig:cartoons} depicts a second possible
configuration.  It consists of a spherically symmetric X-ray emitting region
which is heavily weighted toward the inner wind (in accordance with the LDI
model) and two spiral arms.  As before, the \ha\ emission originates near 
the base of the spiral arms.  To obtain the X-ray variability, we must 
assume that the spiral arms are optically very thick, and that their line of 
sight optical depth decreases with distance from the star.  The latter 
assumption corresponds to the spiral structures expanding with a constant 
solid angle, causing their line of sight column density to decrease as 
$(r/R_\star)^2$ near the star.  As a result, sight lines toward the inner 
wind are strongly obscured by the spiral arms.  When viewed from the right, 
the arms present minimal absorption and 
both the X-ray emission and \ha\ emission are at maximum.  When viewed from 
below, the denser, inner portion of the spiral pattern occults the X-ray 
source and both the \ha\ the X-ray emission are near a minimum.  Finally, 
when viewed from the left, both once again return to a maximum.

This configuration agrees with typical LDI models, since the X-rays are
distributed throughout the wind.  However, in contrast to normal CIR models,
the spiral patterns must have large column densities to account for the
grey variability, and this implies that much of the wind flow is channeled 
through the CIRs.

Regardless of which, if either, configuration is correct, the important
point is that cyclical X-ray variability is inconsistent with our current
understanding of either wind structure formation, X-ray production, or
both.  Consequently, it may provide additional clues on how to interpret
wind structure.  Further, depending on how the wind material is distributed,
the CIRs may account for a small fraction of the wind flow, as predicted by
normal CIR theory \citep{lb08} or, if they originate from regions of 
localized magnetic activity \citep{cb11}, a much larger fraction.
Thus, it is critical to verify our results if we are to arrive at a
self-consistent understanding of stellar winds and mass loss rates in OB 
stars.

\section*{Acknowledgments}
We thank D. Huenemoerder and the CfA \chandra\ support staff for their 
assistance in understanding the data.  We also thank the referee, Anthony 
Moffat, for comments that helped to clarify and enhance the contents of this 
paper.  The scientific results reported in 
this article are based on observations made by the \chandra\ X-ray 
Observatory.  


\label{lastpage}

\end{document}